# Klein tunneling in Weyl semimetals under the influence of magnetic field

Can Yesilyurt,[1] Seng Ghee Tan,[1,2] Gengchiau Liang,[1] and Mansoor B. A. Jalil[1,a)]

[1]*Electrical and Computer Engineering, National University of Singapore, Singapore 117576, Republic of Singapore*

[2]*Data Storage Institute, Agency of Science, Technology and Research (A* Star), Singapore 138634, Republic of Singapore*

Klein tunneling refers to the absence of normal backscattering of electrons even under the case of high potential barriers. At the barrier interface, the perfect matching of electron and hole wavefunctions enables a unit transmission probability for normally incident electrons. It is theoretically and experimentally well understood in two-dimensional relativistic materials such as graphene. Here we investigate the Klein tunneling effect in Weyl semimetals under the influence of magnetic field induced by anti-symmetric ferromagnetic stripes placed at barrier boundaries. Our results show that the resonance of Fermi wave vector at specific barrier lengths gives rise to perfect transmission rings, i.e., three-dimensional analogue of the so-called magic transmission angles in two-dimensional Dirac semimetals. Besides, the transmission profile can be shifted by application of magnetic field, a property which may be utilized in electro-optic applications. When the applied potential is close to the Fermi level, a particular incident vector can be selected for transmission by tuning the applied magnetic field, thus enabling highly selective transmission of electrons in the bulk of Weyl semimetals. Our analytical and numerical calculations obtained by considering Dirac electrons in three regions and using experimentally feasible parameters can pave the way for relativistic tunneling applications in Weyl semimetals.

Klein paradox, which constitutes one of the most interesting consequences of quantum electrodynamics, has been theoretically predicted for massive particles[1-6]. However, the requirement of very high potential drop makes the observation impossible since the potential barrier height must exceed the rest energy of electrons, i.e., $mc^2$ ($m$ being the mass of electron and $c$ the speed of light). The discovery of graphene[7] has enabled the experimental realization of this effect in an accessible condensed matter system, since the effective mass of electrons on graphene is zero, which leads to a zero energy gap between electron and hole states. Therefore, even at low applied potential of < 1 V, electron and hole wavefunctions can be made to match across the barrier, resulting in signature of Klein tunneling, i.e. unit transmission probability at normal incidence[8]. So far, many works have been done to observe the absence of backscattering under an applied potential in graphene[9,10]. In addition, the so-called magic transmission angles caused by the resonance of Fermi wave vector have been investigated in two-dimensional Dirac semimetals such as graphene[11-13]. Additionally, it is important to analyze the Klein tunneling system under magnetic field since only the phase shift in conductance resonances of a ballistic graphene p-n junction may lead a direct evidence of absence of backscattering at the barrier interface[10, 15]. Therefore, it is a prerequisite that any candidate material for Klein tunneling must be analytically treated under the influence of magnetic field as well.

---

a) Electronic mail: elembaj@nus.edu.sg

The recent discoveries[16-18] of material systems exhibiting Weyl semimetallic properties have created a viable avenue for realizing three-dimensional relativistic electron transport. In such materials, the so-called Weyl fermions show gapless energy-momentum relation like graphene, but in all three dimensions in $k$-space. The intriguing nature of Weyl semimetal provides a stable topological state whose low energy dispersion is described by the Weyl Hamiltonian shown in Eq. 1. The Weyl nodes in the bulk of the material always come together in pairs with opposite chirality, carried by the sign of the velocity. This pair of bulk Weyl nodes are connected to one another by the Fermi arc states occurring on the surface. Theoretically, the low energy Hamiltonian of such a system is robust against perturbations since it includes all three Pauli matrices[19]. The exotic features of Weyl semimetals including Weyl nodes in the bulk and Fermi arcs on the surfaces have attracted attention both theoretically and experimentally[20-22]. The surface states (Fermi arcs) have been observed in topological metal[17], TaP[21] and YbMnBi[23] recently. The bulk transport properties have also been investigated, and experimental works have shown consistent agreement with theoretical predictions[23-25]. Moreover, due to the high mobility and chiral nature of electrons in Weyl semimetals, they are expected to be ideal candidates for transport and tunneling applications. Several transport applications such as charge transport[26], magnetotransport[27], extremely large magnetoresistance and ultrahigh mobility[28] have been predicted and observed recently.

In this work, we have investigated the properties of Klein tunneling transmission of Weyl fermions (including the so-called "magic" transmission angles) with and without the influence of magnetic fields, which are induced by four ferromagnetic stripes, as shown in Fig. 1. To obtain the transmission probability ($T$) of the system, we consider electron wavefunctions in the three regions shown, i.e., $\Psi_1$, $\Psi_2$ and $\Psi_3$ for incident, propagated and transmitted electrons, respectively. It is experimentally shown that in three-dimensional Dirac semimetals, the Fermi level as well as potential barrier height, can be tuned by applying a bias to a gated region or by alkali metal doping[29, 30]. By modulating the concentrations of alkali metal dopants in the different regions, one can create Weyl semimetal system with a band-profile as shown in Fig. 1(c). Another alternative would be to apply a back gate to all three regions, and one additional top gate to change carrier concentration only within the barrier region, as was demonstrated in Ref. 30 in thin-film Dirac semimetal $Cd_3As_2$ experimentally. However, the fabricated material must be quite thin (several nanometer thickness) since screening effect is only available for short ranges. As shown in Fig. 1, the application of gated potential or metal doping raises the carrier concentration in the barrier region. If the potential barrier height exceeds the Fermi energy, then the Fermi level will lie within the conduction band in the first and third regions, but cuts across the valence band in the second region. In our scheme, ferromagnetic stripes are placed on top and one side of the Weyl semimetal at the barrier boundaries. The anti-symmetric configuration of ferromagnetic stripes whose magnetic moment perpendicular to the surface induce a localized magnetic field $\boldsymbol{B}$ which is modelled by a delta-function at the boundaries. This translates into a magnetic vector potential $\boldsymbol{A}$ whose spatial profile is shown in Fig. 1(b). The top and side ferromagnetic stripes create a gauge potential on $y$- and $z$-axis respectively. Note that, by controlling the two magnetic fields along $y$- and $z$- direction independently, the effective gauge potential on arbitrary transverse direction can be generated.

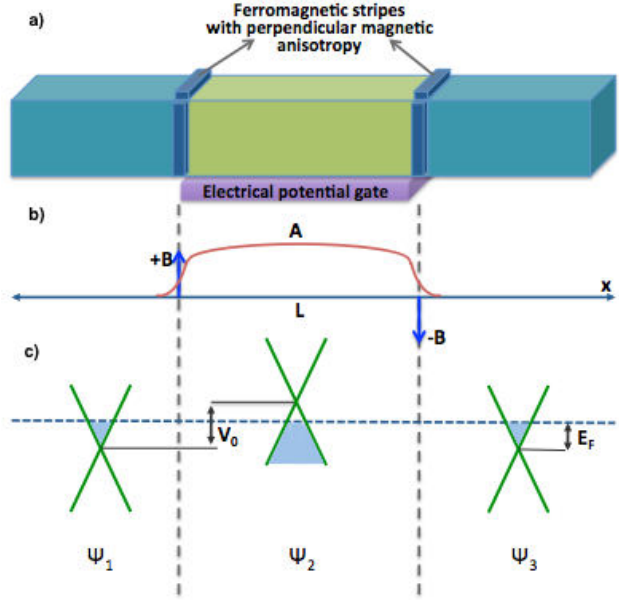

**Fig. 1.** The schematic representation of the single potential barrier Weyl semimetal under the influence of magnetic field induced by four ferromagnetic stripes on the top and one side surfaces. The anti-symmetric anisotropy of ferromagnetic stripes generates a delta function magnetic field **B** thus a magnetic vector potential **A**. Induced potential barrier changes the carrier concentration of barrier region, such that the Fermi level (blue dashed line) cuts across the conduction band in first and third regions, but lies within the valance band in the second region. The occupied states are shown by light blue near the Weyl nodes.

We first consider the case without a magnetic field, for which the low energy Dirac Hamiltonian of the system can be described as follows

$$H = \hbar v_F(\boldsymbol{\sigma}.\boldsymbol{k}) + V_0 = \hbar v_F\big(\sigma_x k_x + \sigma_y k_y + \sigma_z k_z\big) + V_0, \tag{1}$$

where $v_F$ is the Fermi velocity, and $\boldsymbol{\sigma}$ represents all three Pauli matrices. To investigate the transport properties in the bulk system, we consider Weyl electrons near one node and neglect the contribution of surface states (and hence Fermi arcs) to the conduction. By solving the above Hamiltonian, the eigenenergies can be found as $\varepsilon = E_F - V_0 = \pm \hbar v_F k_F$. The positive part of energy represents the electrons carrying negative charge above zero energy while the negative part is for the states that exhibit positively charged behavior (holes) that are unoccupied states in valance band. By considering a transmission of electrons along $x$-direction at angles $\gamma$ (the angle between $\boldsymbol{k}$ and the $xy$ plane) and $\phi$ (the azimuthal angle with respect to the $x$-axis), the components of the eigenstates of equation (1) can be obtained as

$$\psi_{\pm} \equiv \frac{1}{\sqrt{2}} e^{i\boldsymbol{k}\boldsymbol{r}} \begin{pmatrix} 1 \\ e^{i\phi} \sec \gamma \, (\pm 1 + \sin \gamma) \end{pmatrix} \equiv \begin{pmatrix} \psi_a \\ \psi_b^{\pm} \end{pmatrix}.$$

The top component of wavefunction $\psi_a$ for incident, propagating and transmitted regions shown in Fig. 1 can be written as

$$\Psi_{1,a}(x) = e^{ik_x x} + r\, e^{-ik_x x}, \qquad x < 0,$$

$$\Psi_{2,a}(x) = ae^{iq_x x} + b\, e^{-iq_x x}, \qquad 0 < x < L,$$

$$\Psi_{3,a}(x) = te^{ik_x x}. \qquad\qquad x > L.$$

The next step is to write the bottom component of the wave function $\psi_b$ by using the relation $\psi_2 = e^{i\phi} \sec\gamma\,(s + \sin\gamma)\psi_1$, where $s = \pm 1$ depending on the sign of energy $(E_F - V_0)$. Assuming the Fermi energy to be always positive, we have

$$\Psi_{1,b}(x) = e^{i\phi} \sec\gamma\,(1 + \sin\gamma)\, e^{ik_x x} - re^{-i\phi} \sec\gamma\,(1 + \sin\gamma)e^{-ik_x x}; \qquad x < 0,$$

$$\Psi_{2,b}(x) = ae^{i\theta} \sec\alpha\,(s + \sin\alpha)e^{iq_x x} - be^{-i\theta} \sec\alpha\,(s + \sin\alpha)e^{-iq_x x}; \quad 0 < x < L,$$

$$\Psi_{3,b}(x) = t\, e^{i\phi} \sec\gamma\,(1 + \sin\gamma)e^{ik_x x}; \qquad\qquad x > L.$$

In the above, the Fermi wavevector is $k_F = \sqrt{k_x^2 + k_y^2 + k_z^2}$ and the three components of wavevector outside of the barrier can be written as $k_x = k_F \cos\gamma \cos\phi$, $k_y = k_F \cos\gamma \sin\phi$, $k_z = k_F \sin\gamma$, while the $x$-component of wavevector within the barrier is $q_x = \sqrt{\left(\frac{E_F - V_0}{\hbar v_F}\right)^2 - k_y^2 - k_z^2}$. The propagating angles within the barrier $\theta = \tan^{-1}\left(\frac{k_y}{q_x}\right)$ and $\alpha = \tan^{-1}\left(\frac{k_z}{q_x} \cos\theta\right)$ can be found by considering the conservation of the transverse wavevectors $k_y$ and $k_z$ at the barrier interfaces. Finally, the transmission function $t$ can be derived by considering the wavefunction continuity at the barrier interfaces with the proper matching conditions.

In Fig. 2, we analyze the transmission probability $T = |t|^2$ in the case of different applied potential $V_0$ but constant Fermi energy ($E_F \approx 82$ meV). The examples (a) and (c) shows the angular dependence of the tunneling probability when $V_0$ is much greater than $E_F$, which is the case of electron conductance between electron and hole states at the barrier interface. As shown in (b) and (d), the transmission profile of the system is exactly same as the graphene case presented in Ref. 8, if one reduce the system in two-dimension by setting the constant value zero for one of the incident angle ($\gamma = 0$ or $\phi = 0$). The graph (e) shows the case when the $V_0$ is close to $E_F$, thus, the Fermi level is very close to the Weyl node in the barrier. Therefore, only normal incident electrons ($\gamma \approx 0$ and $\phi \approx 0$) can be transmitted perfectly.

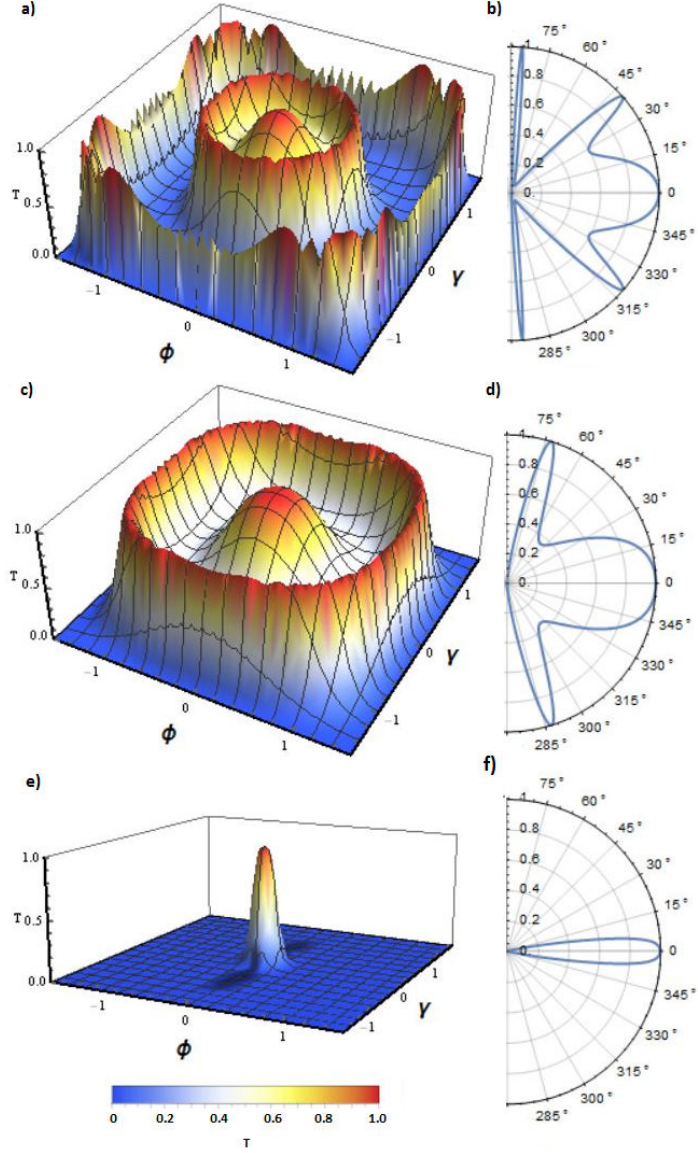

**Fig 2**. Angular dependence of transmission probability for different configuration of applied potentials in the case of $L = 100$ nm and $E_F \approx 82$ meV. (a) and (c) show the cases of $V_0 \approx 200$ meV and 285 meV respectively. (e) shows that when the applied potential is close to Fermi energy ($V_0 = 80$ meV), only normal incidence electrons can transmit as Fermi level is very close to the Weyl node in the barrier region. (b), (d) and (f) represent the cross section of three-dimensional transmissions (a), (c) and (e) in the case of γ=0.

The additional perfect transmission rings can be understood by the resonance condition of Fermi wavevector $k_F$ and the barrier length $L$. These perfect transmission angles can be derived by using the relation $q_x L = \pi n, n = 0, \pm 1, ...$ by substituting the expression of $q_x$ [see Eq. 2]. In two-dimensional relativistic materials such as graphene, the consequence of these resonances enables other perfect transmission angles which satisfy the above condition as shown in Fig. 1 (b), (d) as well as Ref. 8. However, here we found perfect transmission rings instead of points as a consequence of these resonances since two incident angles must satisfy the resonance condition in Weyl semimetal. The number and shape of the perfect transmission rings can be intuitively predicted that they are highly dependent on applied voltage and Fermi energy since $q_x$ is a function of $V_0$ and $E_F$. Also, it can be clearly seen that larger barrier length must enable more perfect transmission rings as shown in Fig 3. Moreover, matching of perfect transmission angles in two different transverse direction results in very interesting tunneling profile when the barrier length L is large enough. This can be understood by the combination of many transmission rings caused by incident angles $\gamma$ and $\phi$. Analytically, one can predict these angles by using resonance conditions and wavevectors as

$$L\sqrt{\frac{(E_F-V_0)^2}{\hbar v_F} - k_F^2 \sin(\gamma)^2 - k_F^2 \cos(\gamma)^2 \operatorname{Sin}(\phi)^2} = \pi\, n, n = 0, \pm 1, \dots \tag{2}$$

Every incident angle that satisfies above expression cause a unit transmission probability ($T = 1$) is depicted by bright pixels in Fig. 3. The predicted transmission profiles shown in Fig. 3 may be experimentally observed by using a point electron source and measuring electron transmission density right after the second interface of the barrier.

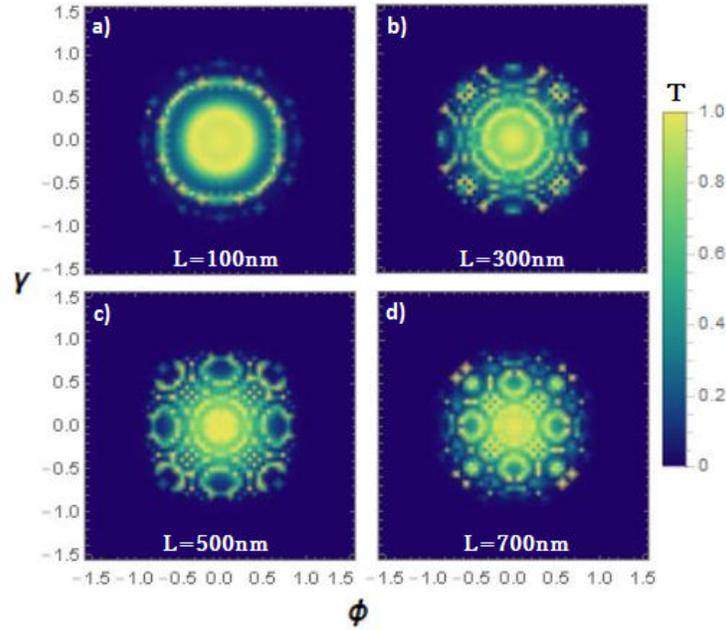

**Fig 3.** Angular dependence of transmission probability in the case of $E_F \approx 82$ meV, $V_0 = 150$ meV for different configurations of the barrier length $L$. Larger barrier length results in more magic transmission rings, and the combination of the perfect incident angles for two transverse directions causes different transmission profiles.

Now, we analyze the transmission profile of Weyl semimetal under the influence of magnetic field induced by ferromagnetic stripes on top and side surfaces. The delta-function magnetic fields at the barrier interfaces cause following gauge potentials on $z$- and $y$-directions.

$$\vec{A}_{Bz} = B_{0(z)}\, l_B [\Theta(x) - \Theta(x - L)]\hat{y}$$

$$\vec{A}_{By} = B_{0(y)}\, l_B [\Theta(x) - \Theta(x - L)]\hat{z}$$

,where $l_B = \sqrt{\frac{\hbar}{|e|B_0}}$. Thus, the Hamiltonian in Equation (1) must be modified as $H = v_F(\boldsymbol{\sigma}.(\boldsymbol{p} + e\boldsymbol{A})) + V_0$. Here, we have neglected the Zeeman splitting since the band shift is very small and only observable under a very high magnetic field of $\approx$ 25 T, as shown experimentally in the Weyl semimetal TaP.[31] Furthermore, it has been shown that at field strength above 3 T, Weyl semimetals like TaP begin to exhibit insulator-like behavior.[31] The magnetic field causes a change on the both transverse wave vectors as $k_y \rightarrow k_y + \frac{eA_{By}}{\hbar}$, $k_z \rightarrow k_z + \frac{eA_{Bz}}{\hbar}$. The wave vector along $y$- and $z$-direction are recalculated with the contribution of the magnetic field; they can be written as

$$k_y = \frac{E_F \cos\gamma \sin\phi + \eta\, v_F \sqrt{|B_{0(y)}|\, \hbar\, |e|}}{\hbar v_F},$$

$$k_z = \frac{E_F \sin\gamma + \eta\, v_F \sqrt{|B_{0(z)}|\, \hbar\, |e|}}{\hbar v_F},$$

where $\eta = \pm 1$ depending on the sign of the magnetic field. Here, the two independent orthogonal magnetic field vectors can create arbitrary angle effective magnetic field that is transverse to the transmission by setting both sign and strength of the magnetic anisotropy of ferromagnetic stripes. Analyzing the angular dependence of the transmission probability, it is shown that electrons are affected by the magnetic field by means of transverse Lorentz displacement. More importantly, a very small range of perfect transmission angles can be selected in two-dimensional transmission space by means of the magnetic gauge potential. Fig. 4 shows the effect of magnetic field on the transmission angles and rings. The perfect transmission rings of Fig. 3(a) are deflected and distorted in the presence of the magnetic field, as shown in Figs. 4(a) and (b). In the case of normal transmission induced by setting an applied potential close to the Fermi energy [see Fig. 2 (e)], the specific wavevector that exhibits absence of backscattering can be tuned by the magnetic gauge potentials. This allows one to generate transmission for a particular incident vector while blocking all other incident vectors. In Figs. 4(c) and (d), the angles corresponding to the absence of backscattering are shifted to different orientations by tuning the magnetic gauge potentials. It can be seen that any point in two-dimensional transmission space can be chosen for unit transmission probability, which may be very useful for electro-optic applications. Moreover, by changing the difference between Fermi energy and barrier height $(E_F - V_0)$, the radius of perfect transmission points can be changed as well.

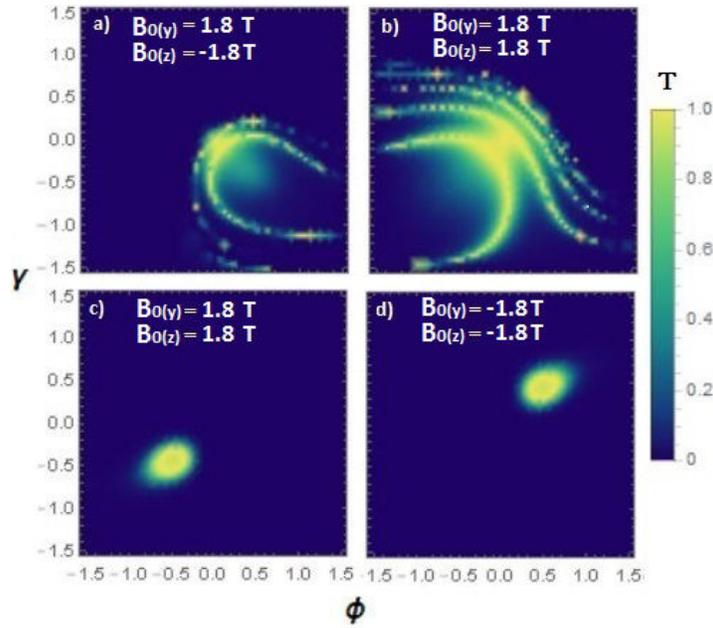

**Fig 3.** Angular dependence of transmission probability under the influence of magnetic field by considering magnetic gauge potentials on $z$- and $y$-component of Fermi wavevector for barrier length $L = 100$ nm. (a) and (b) shows the effect of magnetic field in the case of $V_0 = 140$ meV and $V_0 = 200$ meV respectively. (b) and (c) shows the effect of magnetic field on normal incident electrons when the barrier allows only normal incident electrons since the applied potential is close to Fermi energy ($E_F \approx 82$ meV, $V_0 = 60$ meV). It is seen that application of magnetic field can set the perfect transmission angles on the two-dimensional transmission space.

Thus far, the Klein tunneling has been widely investigated in graphene and successfully realized by direct and indirect measurements[9,10]. Experimental realization of Weyl semimetal has made possible this kind of relativistic tunneling in three-dimension as well. Since the direct evidence of Klein tunneling requires the application of magnetic field to observe the phase shift of conductance oscillations[10], we investigated the tunneling properties and angular dependence of the transmission under the influence of magnetic field with experimentally feasible methods and parameters. The results showed very interesting transmission profiles such as perfect transmission rings due to the resonance conditions. The effect of barrier length on the angular dependence of transmission profile results in an interesting transmission profile due to perfect transmission angles of the combination of two transverse directions. By using a point electron source, these transmission profiles can be experimentally observed by measuring the electron transmission density on the second interface of the potential barrier. In addition, we also analyzed the magic transmission rings in Weyl semimetals as a function of applied potentials and barrier widths. By tunning the potential barrier height, magic transmission rings can be controlled and used for many applications such as electron collimation[31] and Veselago lens[33-35] for electrons in three-dimensional systems. It has been shown previously that the application of asymmetric potential barriers may suppress the magic transmission peaks in graphene[13]. Therefore, further analysis of the magic transmission rings in Weyl semimetals may help realize the transistor effect in such materials. Finally, we showed that the perfect transmission regions can be selected by tuning the magnetic fields which may be used to generate localized transmission in the bulk of Weyl semimetals, a prediction which may be utilized in electro-optic applications.

## Acknowledgments

The authors would like to acknowledge the MOE Tier II grant MOE2013-T2-2-125 (NUS Grant No. R-263-000-B10-112), and the National Research Foundation of Singapore under the CRP Program "Next Generation Spin Torque Memories: From Fundamental Physics to Applications" NRF-CRP9-2013-01 for financial support.